\documentclass[]{article}
\usepackage[preprint]{neurips_2020}
\usepackage[utf8]{inputenc} % allow utf-8 input
\usepackage[T1]{fontenc}    % use 8-bit T1 fonts
\usepackage{hyperref}       % hyperlinks
\usepackage{url}            % simple URL typesetting
\usepackage{booktabs}       % professional-quality tables
\usepackage{amsmath,amsthm, amssymb,amsfonts}       % blackboard math symbols
\usepackage{nicefrac}       % compact symbols for 1/2, etc.
\usepackage{microtype}      % microtypography

\usepackage{algorithmic}
\usepackage{graphicx}
\usepackage{textcomp}
\usepackage{xcolor}
\usepackage[lambda,advantage,operators,sets,adversary,landau,probability,notions,logic,ff,mm,primitives,events,complexity,asymptotics,keys]{cryptocode}
\usepackage{natbib}

\def\BibTeX{{\rm B\kern-.05em{\sc i\kern-.025em b}\kern-.08em
    T\kern-.1667em\lower.7ex\hbox{E}\kern-.125emX}}

\title{
Revisiting Secure Computation Using Functional Encryption: Opportunities and Research Directions\thanks{A conference version of this work is published in \textit{The 2020 Second IEEE International Conference on Trust, Privacy and Security in Intelligent Systems, and Applications}. This is a local version.}
}

\author{%
  Runhua Xu \thanks{Part of work was done before joining IBM Research.}\\
  IBM Research - Almaden \\
  \texttt{runhua@ibm.com} \\
  \And James Joshi \\
  University of Pittsburgh\\
  \texttt{jjoshi@pitt.edu} \\
}

\begin{document}

\maketitle

\thispagestyle{plain}
\pagestyle{plain}

\begin{abstract}
Increasing incidents of security compromises and privacy leakage have raised serious privacy concerns related to cyberspace. Such privacy concerns have been instrumental in the creation of several regulations and acts to restrict the availability and use of privacy-sensitive data.
The secure computation problem, initially and formally introduced as secure two-party computation by Andrew Yao in 1986,  has been the focus of intense research in academia because of its fundamental role in building many of the existing privacy-preserving approaches.
Most of the existing secure computation solutions rely on garbled-circuits and homomorphic encryption techniques to tackle secure computation issues, including efficiency and security guarantees.
However, it is still challenging to adopt these secure computation approaches in emerging compute-intensive and data-intensive applications such as emerging machine learning solutions.
Recently proposed functional encryption scheme has shown its promise as an underlying secure computation foundation in recent privacy-preserving machine learning approaches proposed.
This paper revisits the secure computation problem using emerging and promising functional encryption techniques and presents a comprehensive study.
We first briefly summarize existing conventional secure computation approaches built on garbled-circuits, oblivious transfer, and homomorphic encryption techniques.
Then, we elaborate on the unique characteristics and challenges of emerging functional encryption based secure computation approaches and outline several research directions. 
\end{abstract}

% \begin{IEEEkeywords}
% secure computation, functional encryption, secure aggregation, privacy-preserving machine learning, survey 
% \end{IEEEkeywords}

\section{Introduction}
Devising a well-performing privacy-preserving machine learning (ML) application is crucial due to 
(\romannumeral1) increasing privacy concerns caused by the collection and use of the data, as well as the rapid development and use of AI/Ml and data analytic techniques; and
(\romannumeral2) existing regulations such as the Health Insurance Portability and Accountability Act (HIPPA) and more recent ones such as the European General Data Protection Regulation (GDPR), California Consumer Privacy Act (CCPA), New York SHIELD Act, etc., that have restricted the availability and use of privacy-sensitive data.
Increasingly, privacy-preserving solutions employ various secure computation techniques as the underlying computation building blocks. 

% secure computation - brief intro and its application
% brief intro
Secure computation, also known as secure multi-party computation (SMC), multi-party computation (MPC), or secure function evaluation (SFE), was initially and formally introduced as a general secure two-party computation (2PC),  by Andrew Yao  \citep{yao1982protocols, yao1986generate}.
Work in general secure computation aims to create methods for multiple parties to jointly compute a function over their inputs while keeping those inputs private from each other. 
% its application and examples
With the aim of enabling the utilization of data without compromising privacy, secure computation techniques have been used to solve a wide variety of problems such as privacy-preserving biometric authentication \citep{sadeghi2009efficient, blanton2011secure}, secure search \citep{riazi2017prisearch}, secure auction \citep{kolesnikov2009improved}, and emerging privacy-preserving ML \citep{riazi2018chameleon, gilad2016cryptonets, mohassel2017secureml, rouhani2018deepsecure, xu2019cryptonn, xu2019hybridalpha, chen2019secure}.
Consider a specific example in the healthcare domain.
To find if a person is in a high-risk group for a certain type of cancer, there is a need to compare the person's DNA against a database of cancer patients' DNA.
However, such DNA data is highly privacy sensitive and should not be revealed to other organizations.
Such a dilemma can be solved by using secure multi-party computation that reveals only cancer classification for the targeted person's DNA.

% secure computation - main approaches: gc + ot and he
Since the 1980s, the secure computation problem has been an intensive research topic in academia. Various solutions have been proposed to tackle the issues related to feasibility, efficiency, and security issues in this area. 
Early secure computation approaches rely on boolean circuits, where any function can be represented as a boolean circuit.
Such a circuit can be securely evaluated using the garbled circuits (GC) \citep{yao1986generate} or Goldreich-Micali-Wigderson (GMW) protocol \citep{micali1987play}.
Furthermore, the secure computation components also work with an additional building block, namely, oblivious transfer (OT) that allows a receiving party to obliviously select and receive a message from a group of messages belonging to a sending party without revealing which message was selected. 

Homomorphic encryption (HE) technique that enables computing over the encrypted data by a third-party is another promising candidate to solve the secure computation problem.
In general, there exists two types of methods: \textit{preprocessing model} approach and \textit{pure fully HE} approach.
The SPDZ \citep{damgaard2012multiparty} protocol,  a representative instance of the preprocessing model approach, is built on somewhat homomorphic encryption (SHE) against an active adversary corrupting up to $n$-$1$ of $n$ players. 
To evaluate a function with private inputs, the computation is separated into two phases: an offline preprocessing and an online evaluation phase. 
The pure fully HE approaches directly adopt the fully HE schemes to construct a secure computation protocol.

%combine gc and he
Due to the high computational cost of HE and transmission overhead in GC protocol, the combination of GC and HE techniques also show promise in terms f balancing the computational cost and transmission overhead involved in secure computation solutions.
The first mixed-protocol solution is introduced in \citep{brickell2007privacy}, where HE allows performing \textit{multiplication} and \textit{addition} operations on encrypted values without actually knowing the underlying data, and GC is exploited to implement a secure sub-protocol for the comparison of integer values. 
Then, the TASTY framework \citep{henecka2010tasty} enables the automatic generation of protocols based on GC and HE. 
The hybrid solutions such as ABY \citep{demmler2015aby}, ABY$^3$ \citep{mohassel2018aby3}, and  Chameleon \citep{riazi2018chameleon} propose hybrid solutions by integrating GMW, GC, and additive secret sharing techniques to improve efficiency, scalability, and practicality.

% secure computation - emerging FE
Emerging functional encryption (FE) approach proposed in \citep{lewko2010fully,boneh2011functional} has shown its promise in recently proposed secure computation related privacy-preserving approaches \citep{dufour2018reading, ryffel2019partially, xu2019cryptonn, xu2019hybridalpha, xu2020functional}.
Such FE-based approache0./s present new opportunities to revisit the secure computation problem.
Similar to public-key encryption that allows an owner of a secret key to decrypt any ciphertext created with respect to a corresponding public key, FE is a generalization of public-key encryption in which the decryption key is associated with a function.
Formally, given a ciphertext $\enc_{\pk}(m)$ and a secret key $\sk_{f}$ associated with a function $f$, the holder of $\sk_{f}$ can only learn $f(m)$ and nothing else.
In the last few years, as noted in \citep{abdalla2017multi}, many functional encryption schemes have been proposed. These schemes are aimed towards: (\romannumeral1) creating general functionalities, where the FE constructions are based on more unstable assumptions such as indistinguishable obfuscation or multilinear maps; (\romannumeral2) concrete and efficient realizations for more restricted functionalities, where the functionality mainly covers inner-product and degree-2 polynomials.

The recent adoption of functional encryption schemes of category (\romannumeral2) in the ML domain has shown its promise in terms of communication overhead and computational cost compared to conventional secure computation approaches.
In this paper, we revisit the area of secure computation through use of emerging and promising FE techniques to provide a comprehensive study abd research promise.
Specifically, we first briefly summarize conventional secure computation approaches.
Next, we discuss the unique characteristics and challenges of emerging FE based secure computation approaches. 
We finally outline several directions for future work relevant to a wide range of research communities.

The remainder of this paper is organized as follows.
In Section~\ref{sec:csc}, we introduce previous conventional secure computation solutions built on techniques such as garbled-circuits, oblivious transfer, homomorphic encryption, and secret sharing, etc., and discuss the emerging FE-based secure computation solutions and various related applications in detail in Section~\ref{sec:fe}.
Next, we outline several research challenges and promising future directions in Section~\ref{sec:challenge}.
Finally, we conclude the paper in Section~\ref{sec:conclusion}.

\section{Existing Secure Computation Approaches}
\label{sec:csc}
In this section, we first introduce several fundamental primitives employed for constructing existing conventional secure computation approaches.
Next, we summarize various conventional secure computation approaches built on those fundamental primitives.

\subsection{Foundational Techniques}

\subsubsection{Oblivious Transfer (OT)}
\label{sec:csc:ot}
Oblivious transfer is a cryptographic primitive \citep{rabin2005exchange}. 
% It enables a sender to obliviously transfer one of the information pieces to a receiver without revealing what piece has been transferred to the sender.
It enables a sender to obliviously transfer one of the information pieces to a receiver without revealing to the sender which piece has been transferred. 
Formally, in a $1$-out-of-$2$ oblivious transfer protocol between  Sender ($S$) and Chooser ($C$), denoted as $\mathcal{P}^{\text{OT}^{1}_{2}}_{S,C}$, the input of $C$ is the index $i\in\{0,1\}$ while the inputs of $S$ are two candidate values $x_0, x_1$. At the end of protocol, $C$ only learns $x_i$, while $S$ does not learn which candidate value is chosen.
We only introduce the general idea of the oblivious transfer technique rather than a specific scheme. We refer the readers to \citep{kilian1988founding, rabin2005exchange} for more details.

\subsubsection{Garbled Circuits (GC)}
% \label{sec:csc:gc}
Garbled circuits, initially proposed by Yao \citep{yao1982protocols}, are a fundamental technique in secure multiparty computation. 
The garbled-circuits technique enables secure constant-round computation of any two-party functionality.

Suppose $f_{\mathcal{C}}$ be a boolean circuit that receives two $n$-bit inputs $x,y\in\{0,1\}^{n}$ from Alice ($\mathcal{A}$) and Bob ($\mathcal{B}$), respectively, and outputs $f_{\mathcal{C}}(x,y)\in\{0,1\}$. 
Yao's approach transforms any $f_{\mathcal{C}}$ into the secure garbled circuit $f^{\text{GC}}_{\mathcal{C}}$ and, hence, enables the computation of $f_{\mathcal{C}}(x,y)$ without revealing $x,y$ to each other.
Specifically, for each wire $i$ of $f_{\mathcal{C}}$, $A$ generates two random wire keys $w^{0}_{i}$ and $w^{1}_{i}$ used as labels encoding $0$ and $1$ on that wire, respectively. 

In the garbled-circuits generation, for illustration, let $g$ be a single gate in $f_{\mathcal{C}}$, where input wires to $g$ is labeled as $\mathcal{A}$ and $\mathcal{B}$, and output wire of $g$ is labeled as $\mathcal{O}$, and hence the corresponding wire keys are denoted as $w^{0}_{\mathcal{A}}, w^{1}_{\mathcal{A}}, w^{0}_{\mathcal{B}}, w^{1}_{\mathcal{B}}, w^{0}_{\mathcal{O}}, w^{1}_{\mathcal{O}}$.
The garbled gate $g^{\text{GC}}$ of circuit $f^{\text{GC}}_{\mathcal{C}}$ is defined by a random permutation of the following four ciphertext:
\begin{align*}
    c_{0,0} &= \enc^{\text{SE}}_{w^{0}_{\mathcal{A}}}(\enc^{\text{SE}}_{w^{0}_{\mathcal{B}}}(w^{g(0,0)}_{\mathcal{O}})); \\
    c_{0,1} &= \enc^{\text{SE}}_{w^{0}_{\mathcal{A}}}(\enc^{\text{SE}}_{w^{1}_{\mathcal{B}}}(w^{g(0,1)}_{\mathcal{O}})); \\
    c_{1,0} &= \enc^{\text{SE}}_{w^{1}_{\mathcal{A}}}(\enc^{\text{SE}}_{w^{0}_{\mathcal{B}}}(w^{g(1,0)}_{\mathcal{O}})); \\
    c_{1,1} &= \enc^{\text{SE}}_{w^{1}_{\mathcal{A}}}(\enc^{\text{SE}}_{w^{1}_{\mathcal{B}}}(w^{g(1,1)}_{\mathcal{O}}));
\end{align*}
where $\enc^{\text{SE}}_{w}(x)$ is symmetric key encryption of $x$ using key $w$.
Next, $\mathcal{A}$ garbles all gates of circuit $f^{\text{GC}}_{\mathcal{C}}$ and sends the entire garbled circuit to $\mathcal{B}$.
In the garbled-circuit evaluation phase, 
% still for illustration, 
$\mathcal{A}$ directly sends $w^{b_{\mathcal{A}}}_{\mathcal{A}}$ encoding input bit $b_{\mathcal{A}}$ for its input wire, while launching $\mathcal{P}^{\text{OT}^{1}_{2}}_{\mathcal{A},\mathcal{B}}$ protocol enables $\mathcal{B}$ to learn the wire key $w^{b_{\mathcal{B}}}_{\mathcal{B}}$ without revealing $b_{\mathcal{B}}$ to $\mathcal{A}$.

\subsubsection{Secret Sharing (SS)}
\label{sec:csc:ss}
The secret sharing technique enables distributing a secret among a group of participants, in which each participant is allocated a share of the secret in the sharing phase.
Next, only a sufficient number of shares can reconstruct the secret, while individual shares are of no use on their own.  
In those conventional SMC approaches, two secret sharing schemes are typically adopted: \textit{additive SS} and \textit{t-of-n SS}.
In the \textit{additive SS}, a secret $v$ is shared among $n$ parties such that the addition of those shares yields the secret. 
Conventionally, all operations are performed in a finite field.
Except for the last share, all the rest of the shares are picked randomly in the finite field.
Assume $n$-$1$ of the total $n$ shares are $s_1, s_2, ..., s_{n-1}$, then the final share is computed as $s_n=v-\sum_{i\in[n-1]s_i}$.
In \textit{t-of-n SS}, any $t$ of $n$ shares can be used to recover the secret. The idea is to create a polynomial of degree $t$-$1$ with the secret as the first coefficient and the remaining coefficients chosen at random.
Then, each party receives one of $n$ points on the curve defined by the polynomial. 
In the secret recovery phase, $t$ of $n$ parties collaboratively reveal their points to fit the polynomial of degree $t$-$1$, and hence, the first coefficient, i.e., the original secret, is recovered.

\subsection{Generic SMC}
\label{sec:csc:gc}
Suppose that $f$ is a function to be evaluated securely.
The general process flow of the conventional generic SMC protocol includes five steps: \textit{compiling function}, \textit{garbling circuit}, \textit{transmitting garbled materials}, \textit{oblivious transfer}, and \textit{evaluating circuit}.
For simplicity, we use the two-party setting (i.e., party $\mathcal{P}_{1}$ and $\mathcal{P}_{2}$ collaborate to securely evaluate a function without leaking each one's input) to illustrate each step as follows: 
\begin{itemize}
    \item[(\romannumeral1)]\textit{Compiling Function}: One party, say $\mathcal{P}_{1}$, starts to compile the function $f$. The function $f$ is transformed into a boolean circuit $f_{\mathcal{C}}$ such that $\forall x,y\in \mathcal{X}, \mathcal{Y}, f\in\mathcal{F}, f_{\mathcal{C}}\in\mathcal{F}_{\mathcal{C}}: f(x,y) = f_{\mathcal{C}}(x,y)$, where $\mathcal{X}$ and $\mathcal{Y}$ are the input spaces of two parties, respectively, and $\mathcal{F}$ and $\mathcal{F}_{\mathcal{C}}$ denote the original function space and boolean circuit function space, respectively. 
    As a result, $f_{\mathcal{C}}$ circuit consists of binary gates, where each gate indicates a truth table to compute the gate's output. 
    
    \item[(\romannumeral2)]\textit{Garbling Circuit}: The goal of this step is to garble the truth tables of $f_{\mathcal{C}}$ circuit in the \textit{function compiling} phase.
    Specifically, $\mathcal{P}_{1}$ prepares all garbled boolean circuit gates using the aforementioned garbled circuit technique (as illustrated in Section~\ref{sec:csc:gc}), denoted as $f^{\text{GC}}_{\mathcal{C}}$. As shown in \tablename~\ref{table:garbled_gate}, we use the \textit{AND} gate $g_{\text{AND}}$ as an example to illustrate the \textit{AND} garbled gate $g_{\text{AND}}^{\text{GC}}$.
    
    \item[(\romannumeral3)]\textit{Transmitting Garbled Materials}: In this step, $\mathcal{P}_{1}$ needs to encode its input $x$ into the boolean value $\pmb{b}_{x} \in \{1,0\}^{*}$. Next, $\forall b_i\in\pmb{b}_{x}$, $\mathcal{P}_{1}$ selects corresponding encryption key $w^{b_i}_{i}$.
    Then, $\mathcal{P}_{1}$ sends $\{w^{b_i}_{i}\}_{b_i\in\pmb{b}_{x}}$ and the garbled circuit $f^{\text{GC}}_{\mathcal{C}}$ to $\mathcal{P}_{2}$.
    
    \item[(\romannumeral4)]\textit{Oblivious Transfer}: In the last step, $\mathcal{P}_{2}$ receives the function garbled circuit $f^{\text{GC}}_{\mathcal{C}}$ and keys $\{w^{b_i}_{i}\}_{b_i\in\pmb{b}_{x}}$ corresponding to the input of $\mathcal{P}_{1}$.
    To evaluate the garbled circuit $f^{\text{GC}}_{\mathcal{C}}$, $\mathcal{P}_{2}$ also needs to know the keys $\{w^{b_j}_{j}\}_{b_j\in\pmb{b}_{y}}$ corresponding to its input $\pmb{b}_{y}$; these keys are generated by $\mathcal{P}_{1}$.
    Oblivious transfer technique (as introduced in Section~\ref{sec:csc:ot}) enables 
    $\mathcal{P}_{2}$ to acquire its keys $\{w^{b_j}_{j}\}_{b_j\in\pmb{b}_{y}}$ from $\mathcal{P}_{1}$ without letting $\mathcal{P}_{1}$ learn which keys are chosen.
    
    \item[(\romannumeral5)]\textit{Evaluating Circuit}: With received keys $\{w^{b_i}_{i}\}_{b_i\in\pmb{b}_{x}}$ and $\{w^{b_j}_{j}\}_{b_j\in\pmb{b}_{y}}$,  $\mathcal{P}_{2}$ has to try the decryption for all possible garbled gates; for each gate, only one decryption will work correctly. Then, the result is used to decrypt the next layer of garbled gates in the circuit; this procedure continues through the complete circuit, until finally, $\mathcal{P}_{2}$ can assemble the output bits into the correct output $f(x,y)$ and send back the output to $\mathcal{P}_{1}$.
\end{itemize}

\begin{table}
    \centering
    \footnotesize
    \caption{Example of the AND-garbled table}
    \label{table:garbled_gate}
    \begin{tabular}{cccccc}
        \toprule
            $w_i$ & $w_j$ & $g_{\text{AND}}$ & encrypted output &  & $g_{\text{AND}}^{\text{GC}}$\\
        \midrule
            0 & 0 & 0 & $E_{w^{0}_i, w^{0}_j}(w^{0}_k)$ & $\Rightarrow$ & $E_{w^{1}_i, w^{0}_j}(w^{0}_k)$ \\
            1 & 0 & 0 & $E_{w^{1}_i, w^{0}_j}(w^{0}_k)$ & $\Rightarrow$ &  $E_{w^{0}_i, w^{0}_j}(w^{0}_k)$ \\
            0 & 1 & 0 & $E_{w^{0}_i, w^{1}_j}(w^{0}_k)$ & $\Rightarrow$ &  $E_{w^{1}_i, w^{1}_j}(w^{1}_k)$ \\
            1 & 1 & 1 & $E_{w^{1}_i, w^{1}_j}(w^{1}_k)$ & $\Rightarrow$ &  $E_{w^{0}_i, w^{1}_j}(w^{0}_k)$ \\
        \bottomrule
    \end{tabular}
\end{table}

The aforementioned generic steps are derived from Yao's original approach \citep{yao1982protocols} and only provide basic insight of GC-based secure computation; here, the obvious challenge is the optimization of compiling function to circuits and the transmission overhead.
Severa later studies have focused on tackling efficiency and practicability issues.
The early implementations of generic SMC assumed a semi-honest adversary corrupting a minority of the parties. Typical early work includes Fairplay \citep{malkhi2004fairplay} in two-party setting and its extension FariplayMP \citep{ben2008fairplaymp} in multi-party setting that translates the high-level language called secure function definition language to the boolean circuits, while VIFF \citep{damgaard2009asynchronous} works for arithmetic circuits.
Representative recent GC-based SMC work includes TinyGarble \citep{songhori2015tinygarble} and ObliVM \citep{liu2015oblivm}. TinyGarble framework proposes to generate compact and efficient boolean circuits using industrial logic synthesis tools, where both cyclic and acyclic graph representation of circuits are supported. ObliVM framework provides a domain-specific programming language and a secure computation framework that facilitates the development process.

\subsection{SMC Derived from Homomorphic Encryption}
\label{sec:csc:he}
\textit{Homomorphic Encryption (HE)} is a form of cryptosystem with an additional evaluation capability for computing over ciphertexts without access to the private secret key, in which the result of operations over the ciphertexts, when decrypted, matches the result of operations as if they have been performed on the original plaintext.
Some typical types of HE are \textit{partially} homomorphic, \textit{somewhat} homomorphic, \textit{leveled fully} homomorphic, and \textit{fully} homomorphic encryption according to the the capability of performing different classes of computations.
Unlike traditional encryption scheme that includes three main algorithms: key generation ($\kgen$), encryption ($\enc$), and decryption ($\dec$), an HE scheme also has an extra \textit{evaluation} ($\eval$) algorithm.
Formally, a HE scheme $\mathcal{E}_{\text{HE}}$ includes the above four algorithms such that 
\begin{align*}
    (\pk, \sk) &\leftarrow \mathcal{E}_{\text{HE}}.\kgen(1^{\lambda})\\
    \mathcal{C}_{\text{HE}} &\leftarrow \{\mathcal{E}_{\text{HE}}.\enc_{\pk}(m_1), ..., \mathcal{E}_{\text{HE}}.\enc_{\pk}(m_n)\} \\
    \mathcal{C}^{f}_{\text{HE}} &\leftarrow \mathcal{E}_{\text{HE}}.\eval_{\pk}(f,\mathcal{C}_{\text{HE}}) \\
    f(m_1, ..., m_n) &\leftarrow
    \mathcal{E}_{\text{HE}}.\dec_{\sk}(\mathcal{C}^{f}_{\text{HE}})
\end{align*}
where $\{m_1,...,m_n\}$ are the message to be protected, $\pk$ and $\sk$ are the key pairs generated by the key generation algorithm.

Since the notion of HE has been proposed, several practical implementations have been released to promote its adoption in various scenarios, including secure outsourcing computation and generic secure multiparty computation. 
The Paillier cryptosystem \citep{paillier1999public} is an additive partially homomorphic encryption system, where given the message $m_i$ and $m_j$, the Paillier system $\mathcal{E}_{\text{HE}}^{\text{Paillier}}$ satisfies the HE equation such that 
\begin{align*}
    \mathcal{E}_{\text{HE}}^{\text{Paillier}}.\enc(m_i)\cdot\mathcal{E}_{\text{HE}}^{\text{Paillier}}.\enc(m_j)=\mathcal{E}_{\text{HE}}^{\text{Paillier}}.\enc(m_i+m_j)
\end{align*}
The HElib \citep{halevi2014algorithms} implemented several typical fully HE schemes with applied optimization techniques like bootstrapping, smart-vercauteren, and approximate number.
SEAL \citep{sealcrypto2020} is another HE library that allows additions and multiplications to be performed on encrypted integers or real numbers. Other operations, such as encrypted comparison, sorting, or regular expressions, in most cases, are not feasible to evaluate on encrypted data using this technology. 

% adoption in SMC
To achieve generic SMC using HE technique two representative approaches are usually available: \textit{preprocessing model} approach and \textit{pure fully homomorphic encryption} (FHE) approach.
In the former approach, a trusted dealer is assumed, where the dealer does not need to know the function to be computed, nor the inputs, but it provides raw materials for the computation. The dealer can be implemented by a secure protocol using public-key techniques. These operations can be run in a preprocessing manner based on \textit{somewhat homomorphic encryption} (SHE) schemes.
Next, the \textit{online} protocol uses only inexpensive information-theoretic primitives to securely evaluate a function.
The \textit{pure FHE} approach, derived with the approach of FHE by Gentry \citep{gentry2009fully}, is more straightforward than the \textit{preprocessing model} approach, where all the parties first encrypt their input under the FHE scheme and evaluate the desired function on the ciphertexts using the homomorphic properties.
Next, these parties can perform a distributed decryption on the final texts to get the results. 
The apparent advantage of the \textit{pure FHE} approach is communication efficiency (i.e., low bandwidth consumption); however, it comes at a price that existing FHE is not computationally efficient and then can only evaluate small circuits.
The \textit{preprocessing model} approach essentially compromises the design, where it needs more communication and number of rounds, but the computational overhead is much smaller.

\subsection{Achieving SMC in Hybrid Manner}
The mixed-protocols solution that combines the techniques mentioned above is another direction to achieve efficient and practical secure multi-party computation. 
The general idea of those mixed-protocol approaches is to evaluate operations according to their best efficient representations.
The additions and multiplications with an efficient representation as an arithmetic circuit can use a homomorphic encryption approach. In contrast, the comparisons with an efficient representation as a boolean circuit will use Yao's garbled circuits technique.

Some representative mixed-protocols solutions include TASTY \citep{henecka2010tasty}, ABY \citep{demmler2015aby}, ABY$^3$ \citep{mohassel2018aby3}, Chameleon \citep{riazi2018chameleon}, etc.
TASTY \citep{henecka2010tasty} is a compiler that can generate protocols based on HE, efficient garbled circuits, and their combinations.
ABY \citep{demmler2015aby} is a mixed-protocol framework that efficiently combines secure computation schemes based on arithmetic sharing, boolean sharing, and garbled circuits, in which all cryptographic operations are pre-computed and then provides efficient conversions among secure computation schemes based on pre-computed oblivious transfer extensions.
Next, ABY$^3$ \citep{mohassel2018aby3} is proposed by extending the ABY framework in the three-party setting for privacy-preserving ML.
To improve the performance in terms of computation and communication between parties, the framework Chameleon \citep{riazi2018chameleon} that is also based on ABY overcomes two limitations such as adopting a semi-honest third-party for preprocessing arithmetic triples instead of oblivious transfer used in ABY and handling signed fixed-point numbers.

\section{Secure Computation Using Emerging Functional Encryption}
\label{sec:fe}
Since the notion of emerging functional encryption (FE) was introduced in \citep{lewko2010fully,boneh2011functional}, FE has attracted increasing attention and interests. 
Similar to HE design, FE also allows evaluating a function over the encrypted dataset, but it is different in terms of key management. 
The keys (i.e., public and private keys) of HE are generated in pairs. In contrast, keys (i.e., public and function derived keys) in FE are generated separately by a trusted third-party authority that holds a master secret key. 
Furthermore, another difference between FE and HE is that given an arbitrary function $f(\cdot)$, HE allows a third-party facility with issued public key to compute \textit{an encrypted result of} $f(x)$ over an encrypted $x$, whereas FE allows the third-party facility with issued function derived key to compute \textit{a plaintext result of} $f(x)$ from an encrypted $x$ \citep{alwen2013relationship}.
Intuitively, the function computation party in the HE (i.e., the evaluation party) can only contribute its computation resource to obtain the encrypted function result but cannot learn the function result unless it has the secret key. In contrast, the party computing function in the FE scheme (i.e., usually, the decryption party) can obtain the function result with the issued function derived key.

In the following subsections, we first introduce and discuss various representative FE schemes.
Next, we introduce several existing secure computation approaches based on those FE constructions and their potential applications.

\subsection{Functional Encryption Overview}
FE is a generalization of public-key encryption in which one party with an issued function derived secret key is allowed to only learn a function of what the ciphertext is encrypting.
Formally, FE $\mathcal{E}_{\text{FE}}$ includes four algorithms: \textit{setup} ($\mathsf{Setup}$), \textit{key generation} ($\kgen$), \textit{encryption} ($\enc$) and \textit{decryption} ($\dec$) algorithms such that
\begin{align*}
    (\pk,\mathsf{msk}) &\leftarrow \mathsf{Setup}, \\
    (\sk_{f}) &\leftarrow \kgen(f, \mathsf{msk}), \\
    \mathcal{C}_{\text{FE}} &\leftarrow \{\mathcal{E}_{\text{FE}}.\enc_{\pk}(m_1), ..., \mathcal{E}_{\text{FE}}.\enc_{\pk}(m_n)\}, \\
    f(m_1, ..., m_n) &\leftarrow \mathcal{E}_{\text{FE}}.\dec_{\sk_{f}}(\mathcal{C}_{\text{FE}}),
\end{align*}
where $\{m_1,...,m_n\}$ are the messages to be protected; the $\mathsf{Setup}$ algorithm creates a public key $\pk$ and a master secret key $\mathsf{msk}$, and $\kgen$ algorithm uses $\mathsf{msk}$ to generate a new functional private key $\sk_f$ associate with the functionality $f$.

Except for several recently proposed decentralized FE schemes \citep{chotard2018decentralized,abdalla2019decentralizing,chotard2020dynamic}, the classical FE schemes rely on a trusted third-party authority to provide key service, such as issuing a functional private key associated with a specific functionality via $\kgen$ algorithm.
In general, existing FE schemes research can be broadly categorized as focusing on (\romannumeral1) feasibility for general functionalities and (\romannumeral2) concrete, efficient, and practical realizations for restricted functionalities.
%limitation in summary
Existing constructions of type (\romannumeral1) FE schemes are all inefficient or impractical as these constructions rely on quite unstable assumptions such as indistinguishability obfuscation or impose severe restrictions on the number of issued secret keys. In contrast, constructions of type (\romannumeral2) are known only for the case of linear and quadratic functions.

\subsubsection{FE for Generic Functionality}
The goal of type (\romannumeral1) FE schemes is to support more general functionalities. 
Typical constructions include the work proposed in \citep{goldwasser2014multi, ananth2015selective, brakerski2018multi}, but unfortunately, they all rely on non-standard cryptographic assumptions such as indistinguishability obfuscation, single-input FE for circuits, or multilinear maps.
Specifically, the problem of multi-input FE is studied in \citep{goldwasser2014multi, ananth2015selective}, where the construction is based on the indistinguishability obfuscation primitive.
% In the multi-input FE, a secret key can correspond to an n-ary function that takes multiple ciphertexts as input.
Beyond the indistinguishability obfuscation, the work proposed in \citep{brakerski2018multi} provides various assumptions such as multilinear maps and single-input FE for circuits.

\subsubsection{FE for Restricted Functionality}
Few recent schemes of type (\romannumeral2) such as those proposed in \citep{abdalla2015simple, abdalla2017multi, abdalla2019decentralizing, chotard2018decentralized, baltico2017practical, gay2020new} that focus on some more restricted functionalities, e.g., inner-product $\mathcal{F}_{\text{IP}}$ and quadratic polynomials $\mathcal{F}_{\text{QP}}$, but in a more efficient and practical way, where the construction uses the standard assumptions such as decisional Diffie-Hellman (DDH) and learning with errors (LWE).
Take the inner-product FE schemes as an example; each function $f_{\text{IP}} \in \mathcal{F}_{\text{IP}}$ is specified by a collection $\pmb{y}$ of $n$ vectors $\{\pmb{y}_1, ..., \pmb{y}_n\}$ and takes a collection of $n$ vectors $\{\pmb{x}_1, ..., \pmb{x}_n\}$ as input and produces $f_{\text{IP}}=\sum_{i\in n}\langle\pmb{x}_i,\pmb{y}_i\rangle$ as output.

\subsubsection{FE Scheme with Additional Features}

Accompanying with the practical FE construction for restricted functionalities, most recent works tries to add more features to existing FE schemes such as \textit{labeled} FE \citep{chotard2018decentralized}, \textit{decentralized} FE \citep{abdalla2019decentralizing}, \textit{dynamic decentralized} FE \citep{chotard2020dynamic}, and FE with access control \citep{abdalla2020inner}.
The labeled FE tries to fill the gap between the multi-input and multi-client scenarios. Every ciphertext for every slot can be combined with any other ciphertext for any other slot in multi-input FE. Hence, the function derived keys are allowed to decrypt an exponential number of values when various ciphertexts from each party are permuted and combined.  
In multi-client FE, there is a label involved in the encryption phase; only the ciphertext encrypted under the same label can be successfully decrypted using the function derived key. Hence, each client has more control over how much information is leaked about their data.
Most of the FE schemes mentioned above rely on a third-party authority (TPA) that holds a master secret key to provide key service to the decrypting party.
The goal of decentralized FE is to remove the central TPA to make the FE scheme better suited for real-word scenarios, where the function derived key is generated by the collaboration of encrypting parties. 
The dynamic FE focuses on the case of the dynamic group of encrypting parties using decentralized FE.

\subsection{Secure Computation Through FE}
Except for the FE schemes for generic functionality, some recent FE schemes for restricted functionalities have shown promise to build a variety of practical secure computation protocols that enable more complex privacy-preserving ML such as emerging federated learning and deep learning. 

To illustrate the core idea of various FE-based secure computation protocols, suppose that there exists one coordinator $\mathcal{C}$ and a set of multiple data sources/clients $\{\mathcal{D}_1, ..., \mathcal{D}_{n}\}$, where each client $\mathcal{D}_{i}$ has its input $\pmb{x}_{i}$.
Most of the existing FE-based secure computation approaches have the following threat models and assumptions: (\romannumeral1) the coordinator is assumed to be \textit{honest-but-curious}, that indicates the coordinator may try to infer the information from the data sources but strictly follow the instructions/protocols; (\romannumeral2) each data source could be benign or malicious, and the malicious ones could not infer the information from the benign data sources; (\romannumeral3) the third-party authority (TPA) that provides key service is assumed to be fully trusted.
Note that the TPA is an optional entity in emerging FE-based secure computation protocols as decentralized FE has shown its promise to support fewer restricted functionalities.  

\subsubsection{Secure Two-party Computation}
\begin{figure}
    \centering
    \includegraphics[scale=0.45]{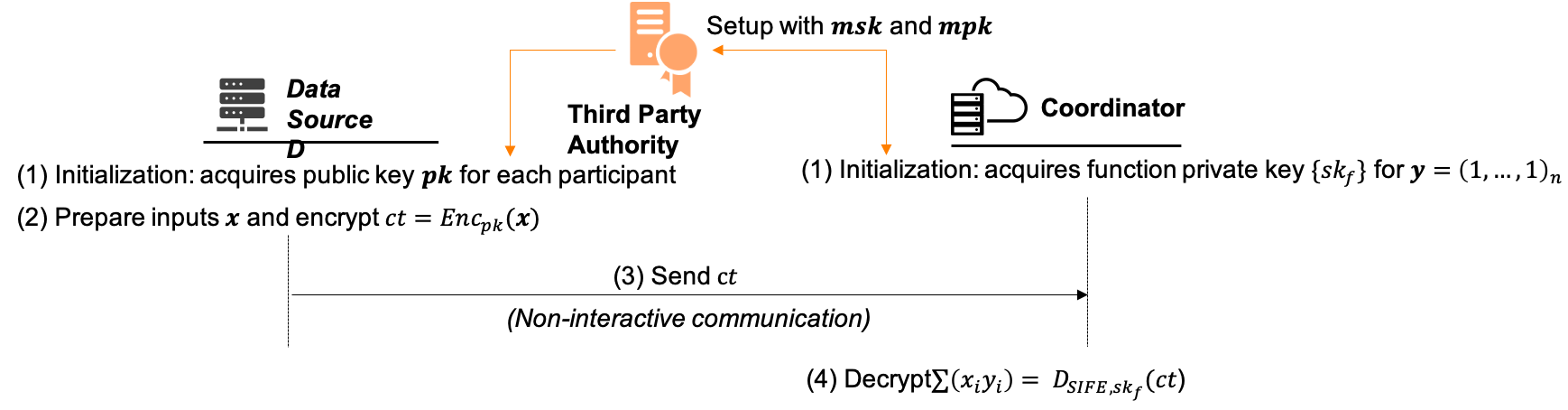}
    \caption{Secure two-party computation protocol using FE-based approach}
    \label{fig:s2pc}
\end{figure}
In the secure two-party computation scenario, coordinator $\mathcal{C}$ plays the role of data provider that takes as input $\pmb{y}$ and securely evaluates the function $f(\pmb{x},\pmb{y})$ without learning the input data from the data source $\mathcal{D}$.
As depicted in \figurename~\ref{fig:s2pc}, we illustrate the general approach to adopting the FE schemes in two-party computation protocols.
It includes three phases:
\begin{itemize}
    \item[(\romannumeral1)] \textit{Protocol Initialization}. The protocol is initialized by setting up public key $\pk$ for the data source and issuing function derived key $\sk_{f}$ for the coordinator. 
    
    \item[(\romannumeral2)] \textit{Data Protection}. The data source employs the encryption algorithm to protect its input data $\pmb{x}$ using the adopted FE scheme and sends the ciphertext to the coordinator for function evaluation. 
    
    \item[(\romannumeral3)] \textit{Ciphertext Computation}. The function evaluation is more straightforward. The coordinator decrypts the received ciphertext, and the decryption result is the function evaluation result.
\end{itemize}
Unlike the garbled-circuit-based SMC, the secure function evaluation procedure is implied in the decryption procedure. 
Compared to the HE based SMC that separates the function evaluation procedure and function result release procedure into different entities, the FE-based secure computation approach needs to release the function result to the decryption party. Such a design is similar to the setting in garbled-circuit-based SMC approaches.

\subsubsection{Secure Multi-party Computation}

\begin{figure*}
    \centering
    \includegraphics[scale=0.45]{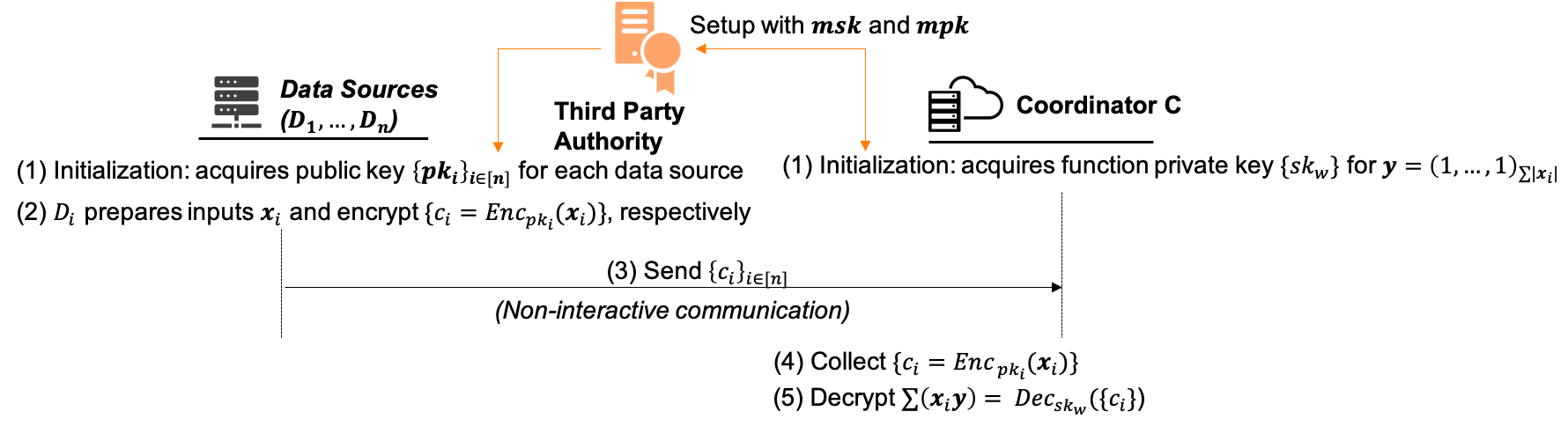}
    \caption{Secure multi-party computation protocol using FE-based approach}
    \label{fig:smpc}
\end{figure*}

In secure multi-party computation scenario, $\mathcal{C}$ can coordinate the secure evaluation among a set of data sources $\{\mathcal{D}_1, ..., \mathcal{D}_{n}\}$ as the role of the coordinator,
It can also take its data as part of inputs for the evaluated function and hence play the role of a participant instead of the role of a coordinator.
The supported secure evaluation functions include secure aggregation, weighted secure aggregation, quadratic function, etc..
Similar to FE-based secure two-party computation setting, as shown in \figurename~\ref{fig:smpc}, the secure multi-party computation protocol also includes three phrases:
\begin{itemize}
    \item[(\romannumeral1)] \textit{Protocol Initialization}. The protocol is initialized by setting up public key $\pk_{\mathcal{C}_i}$ for each data source $\mathcal{C}_i$ and issuied function derived key $\sk_{f}$ for the coordinator. 
    
    \item[(\romannumeral2)] \textit{Data Protection}. Each data source $\mathcal{C}_i$ employs the encryption algorithm of FE scheme with issued party-specific key $\pk_{\mathcal{C}_i}$ to protect its input data $\pmb{x}_i$ and sends the ciphertext to the coordinator for function evaluation. 
    
    \item[(\romannumeral3)] \textit{Ciphertext Computation}. The coordinator first collects the ciphertext from each data source and then executes the decryption process over the ciphertext set directly. The decryption result is the function evaluation result.
\end{itemize}
Note that according to different roles of the coordinator, there exist different settings:
(\romannumeral1) in the case that $\mathcal{C}$ is enrolled as a data contributor, the setting is similar to that in the secure two-party computation, $\mathcal{C}$ only needs to request its function derived key $\sk_{f}$ by providing its input, e.g., $\pmb{x}_{\mathcal{C}}$, and then acquires $f(\{\pmb{x}_i\}, \pmb{x}_{\mathcal{C}})$ by launching the decryption process;
(\romannumeral2) in the case that $\mathcal{C}$ only plays the role of the coordinator, $\mathcal{C}$ can request its function derived key $\sk_{f}$ by providing a (weight) vector $\pmb{w}$, to measure the weight of each data source's input. 
For instance, for the simple average secure aggregation scenario, $\pmb{w}$ can be set as an all-one vector.

\noindent\textit{Remark}.
According to the aforementioned FE-based secure computation architecture, the supported underlying functions of secure computation protocols rely on the implicitly supported functionality of the underlying FE scheme. 
Thus, recently proposed FE-based secure computation protocols only support a limited number of secure computation scenarios because of the limited functionalities supported by existing efficient and practical FE schemes. 
However, those protocols have shown their promise from the perspective of practicality and efficiency compared to state-of-the-art secure computation approaches constructed from other techniques, as discussed in Section~\ref{sec:csc}.
In the next subsection, we introduce several useful applications that employ FE-based secure computation approaches discussed here.

\subsubsection{Applications of FE-based secure computation}
\label{sec:fe:app}

Here, we introduce emerging privacy-preserving applications proposed in \citep{baltico2017practical, dufour2018reading, ryffel2019partially, xu2019cryptonn, xu2019hybridalpha, xu2020functional} that adopt the FE-based secure computation to support the need for privacy protection in the computation phase.

A fundamental and straightforward privacy-preserving application is \textit{(weighted) secure aggregation}. The problem of computing a sum from different parties' inputs where none of the parties reveals its input in the clear - even to the coordinator in some architectures - is referred to as \textit{secure aggregation}.
To solve the secure aggregation problem, various approaches have been proposed in the literature including (\romannumeral1) conventional secure computation approaches using garbled circuits techniques and partially/fully HE as introduced in Section~\ref{sec:csc}, (\romannumeral2) approaches based on anonymous communication using mix-nets \citep{chaum1981untraceable} or DC-nets\citep{chaum1988dining}, and (\romannumeral3) approaches based on pairwise masking \citep{bonawitz2017practical,ion2019deploying}.
As discussed above, approach (\romannumeral1) has its limitation due to the high computation cost of HE and transmission overhead in garbled-circuit techniques.  
In approach (\romannumeral2), anonymous communication only protects the links between the privacy-sensitive data and the participants, and hence, these techniques do not prevent private information from leaking.
Furthermore, the system setup and communication complexity in approach (\romannumeral3) are not as practical and efficient as expected compared to the emerging FE-based secure aggregation approach.

Federated learning (FL) has been recently proposed to address privacy problems by allowing collaborative training of machine learning models among parties where each party can locally hold its data.
However, this ML training approach still poses privacy risks such as inference attacks \citep{nasr2019comprehensive, shokri2017membership}.
To address such privacy leakage, several techniques, including secure computation, have been adopted to achieve privacy-preserving FL (PPFL).
For example, some representative PPFL systems include the works proposed in \citep{pettai2015combining,truex2019hybrid,bonawitz2017practical} that combine differential privacy techniques and secure aggregation techniques.
The experimental evaluation reported in \citep{xu2019hybridalpha} indicates that the FE-based secure aggregation show significant improvement in terms of practicality, computation cost, and transmission overhead compared to existing conventional secure computation approaches \citep{xu2019hybridalpha}. 

Privacy-preserving deep learning (PPDL) is another type of privacy-preserving application that builds on secure computation techniques to protect training data privacy leakage while still generating a well-trained model. 
For instance, various secure computation approaches have been proposed in \citep{rouhani2018deepsecure, xu2019cryptonn, nandakumar2019towards, gilad2016cryptonets} to achieve PPDL.
Except for the garbled-circuit-based PPDL solutions \citep{rouhani2018deepsecure}, as reported in \citep{xu2019cryptonn, xu2020functional}, FE-based PPDL has also shown its efficiency promise in training time without making a compromise on model accuracy compared to the conventional HE based PPDL solutions as illustrated in \citep{nandakumar2019towards, gilad2016cryptonets}.

\section{Challenges and Future Directions}
\label{sec:challenge}

Even though the notion of secure computation was proposed in the 1980s, it is still an active and ongoing research area. 
The emerging FE technique has shown to be a promising direction to achieve secure computation beyond the conventional secure computation approaches that we have introduced and discuss in Section~\ref{sec:fe}.
This section briefly outlines key promising research directions for FE-based secure computation.

\subsection{Enriching Functionality}
The successful adoption of FE-based secure computation in PPFL and PPDL applications, as illustrated in Section~\ref{sec:fe:app}, is based on FE schemes for limited functionalities.
For instance, the functionality of the FE scheme adopted in \citep{xu2019cryptonn, xu2019hybridalpha} is only limited to the inner-product. 
There is a lack of FE schemes to support more functionalities such as comparison and max/min operations, degree-n polynomial computation.
Meanwhile, these FE schemes that support various functionalities aforementioned can still be constructed in an easy and efficient approach. 
It is important to enrich FE schemes' with to support more functions so that we can construct more function-specific secure computation protocols such as trigonometric, exponentials and logarithmic functions to support more privacy-preserving application scenarios. 

\subsection{Enhancing Security and Privacy Guarantees}
Security and privacy guarantees are critical issues in any secure computation protocols. 
The security strength of FE-based secure computation relies on the security of underlying FE schemes. Existing FE schemes with practical constructions usually against the selective indistinguishability under chosen-plaintext attack (IND-CPA) in realizable security assumptions such as the decisional Diffie–Hellman (DDH) assumption.
Even though such FE constructions satisfy security requirement in most scenarios, it may not meet the application scenarios, such as military-related applications, where stricter security requirements/guarantees need to be ensured. Further, as we look ahead, stronger constructions need to be explored to provide protection against an adversary using quantum computing. 
As the function result is released to the assumed \textit{honest-but-curious} coordinator, as illustrated in Section~\ref{sec:fe}, there may exist potential private information leakage, as reported in \citep{xu2019hybridalpha}.
% Thus, there is still a lack of study to enhance the FE schemes' security and privacy guarantee of FE-based secure computation protocols. 
Thus, there is still a a need to explore stronger security and privacy guarantees for FE-based secure computation protocols, including for post-quantum era.

\subsection{Increasing Efficiency}
Efficiency issue has bee the primary objective in SC area since the first secure two-party computation protocol was proposed. 
The main effort of years of research has been to pursue efficiency of SC so that they can be practically deployed. 
For instance, in the garbled-circuit based SC, the recent efforts still aim at improving the efficiency by lowering the communication payload and increasing the compilation efficiency. In contrast, the HE-based secure computation focuses on the function evaluation efficiency over the ciphertext, and decryption time.
Even though FE-based secure computation approaches have shown its promise in efficiency improvements compared to the HE-based approach, it is still a challenge to adopt FE-based secure computation in large-scale private data analysis. Hence, exploring more improved the efficient techniques for FE-based secure computation is essential direction.

\subsection{Dynamic, Decentralized, and Threshold FE-based Secure Computation}
As illustrated in PPDL and PPFL using FE-based secure computation \citep{xu2019cryptonn, xu2019hybridalpha}, the underlying FE schemes rely on a third-party authority (TPA) to provide key service, where the TPA is assumed to be fully trusted in their threat models. 
Decentralized FE schemes that do not rely on a TPA makes the FE-based secure computation more applicable in a real scenario, where there is a challenge to deploy the TPA.
Furthermore, the feature of supporting the dynamic participant group is also an important research direction to make the secure computation protocol resistant to an unexpected drop-off of some participants without impacting the current execution of a protocol.
Lastly, extra features such as threshold decryption and access control on encrypted data could also be possible research directions.
For instance, there is need to explore devise FE-based secure computation protocols to support scenarios that encompass integrated IoT-edge-cloud computing environments. 
Furthermore, the data owners should have full control of how and when to release the function results over their encrypted data - thus, user-centric approaches needs to be explored.

\subsection{Privacy-Preserving Applications}
Emerging FE-based privacy-preserving approach proposed in \citep{baltico2017practical, dufour2018reading, ryffel2019partially, xu2019cryptonn, xu2019hybridalpha, xu2020functional} have shown promise for using FE-based approaches for privacy-preserving ML. 
There is a number of complex and challenging problems in terms of secure computation to be tackled in the privacy-preserving ML applications.  
For instance, how to extend the FE-based secure computation in federated learning from horizontal setting to vertical setting; how to support FE-based secure computation in more types of neural networks such as recurrent neural network and transformer in privacy-preserving deep learning; how to use FE-based secure computation in more types of algorithms such as decision tree, ensemble model, Bayesian related models, cluster algorithm, etc..
Supporting complex privacy-preserving applications built on existing simple FE-based secure computation protocols is still a considerable challenge.

\subsection{Transparent and Accountable Crypto Infrastructure}
Except for the decentralized FE schemes proposed, most FE schemes, as well as other emerging cryptographic schemes such as attribute-based encryption and predicate encryption, rely on a trusted third-party authority (TPA) to provide key service.
However, unlike the widely deployed certificate authority (CA) \citep{li2019certificate} infrastructure, there is a lack of widely trusted TPA infrastructure to support those TPA-related FE-based secure computation protocols in the Internet environment. 
With the widely trusted and transparent TPA infrastructure \citep{xu2020trustworthy}, it can simplify the deployment of FE-based privacy-preserving applications in the real Internet environment and accelerate the use of privacy-preserving applications. Such a infrastructure for transparency and accountability will ensure more trustworthy deployment of FE-based SC mechanisms.

\subsection{Realization and Open Source Library}
Unlike the HE based secure computation approaches where there exist several open-source libraries such as HElib from IBM \citep{halevi2014algorithms} and Microsoft SEAL \citep{sealcrypto2020}, there is still a lack of standard or widely used library for FE schemes.
The HElib \citep{halevi2014algorithms} implement several typical fully HE schemes with applied optimization techniques like bootstrapping, smart-vercauteren, and approximate number.
SEAL \citep{sealcrypto2020} is another HE library that allows additions and multiplications to be performed on encrypted integers or real numbers. 
Similar to that for HE open source tools, there is a need to establish open-source practical FE libraries that can promote and accelerate the deployment of of FE-based secure computation protocols for privacy-preserving solutions.

\subsection{Benchmarks}
Finally, as FE-based secure computation is a nascent research area, we are at a pivotal time to shape the developments in this area.
It is critical for the broader researchers from other communities such as ML to identify which FE-based secure computation solutions satisfy their privacy protection needs when considering the impact of efficiency, scalability, and supported functionality. 
Besides, there is also a lack of comparitive evaluation of all possible secure computation approaches under the same or similar experimental conditions.
Thus, it is important to establish benchmarks for existing FE-based secure computation approaches, even for all other secure computation approaches, to help support indepth evaluations if newly proposed solutions.

\section{Conclusion}
\label{sec:conclusion}
In this article, we revisited secure computation using emerging functional encryption techniques and discussed their promise to enhance secure computation solutions.
We have provided an overview of conventional secure computation protocols/approaches and their corresponding fundamental components and primitives.
We have discussed the unique characteristics of FE-based secure computation approaches and compared them to other conventional secure computation solutions.  
Finally, we have discussed the associated challenges of FE-based secure computation approaches and outlined open problems and future research directions.

\section*{Acknowledgment}
This work was performed while James Joshi was serving as a Program Director at NSF; and the work represents the authors' views and not that of NSF's.

\bibliographystyle{plainnat}
\bibliography{reference}

\end{document}